\documentclass[a4paper]{article}

\usepackage{INTERSPEECH2020}
\usepackage{enumitem}
\usepackage{float}

\title{Detecting Distrust Towards the Skills of a Virtual Assistant Using Speech}

\name{Leonardo Pepino$^{1,2}$, Pablo Riera$^2$, Lara Gauder$^{1,2}$, Agust\'in Gravano$^3$, Luciana Ferrer$^2$,}
\address{
$^1$Departamento de Computaci\'on, FCEyN, Universidad de Buenos Aires (UBA), Argentina\\
$^2$Instituto de Investigaci\'on en Ciencias de la Computaci\'on (ICC), CONICET-UBA, Argentina\\
$^3$Escuela de Negocios, Universidad Torcuato Di Tella, Argentina}
\email{\{lpepino,priera,mgauder,lferrer\}@dc.uba.ar; agravano@utdt.edu}

\begin{document}

\maketitle
\begin{abstract}
Research has shown that trust is an essential aspect of human-computer interaction directly determining the degree to which the person is willing to use the system. An automatic prediction of the level of trust that a user has on a certain system could be used to attempt to correct potential distrust by having the system take relevant actions like, for example, explaining its actions more thoroughly.
In this work, we explore the feasibility of automatically detecting the level of trust that a user has on a virtual assistant (VA) based on their speech. We use a dataset collected for this purpose, containing human-computer speech interactions where subjects were asked to answer various factual questions with the help of a virtual assistant, which they were led to believe was either very reliable or unreliable. We find that the subject's speech can be used to detect which type of VA they were using, which could be considered a proxy for the user's trust toward the VA's abilities, with an accuracy up to 76\%, compared to a random baseline of 50\%. These results are obtained using features that have been previously found useful for detecting speech directed to infants and non-native speakers.
\end{abstract}

\section{Introduction}

Just as trust is an important factor in human communication, it is also considered an essential part of human-computer interactions \cite{parasuraman1997,drnec2016trust}. Too much or too little trust in a system can cause the user to over-use or under-use its capabilities, respectively. Ideally, a system should be able to track the user's level of trust, which would allow it to act accordingly, attempting to calibrate their trust to an ideal level \cite{muir1994,okamura2020,drnec2016trust}. 

For systems that communicate with the user through speech, one possible way to track the level of trust is through the user's voice. It is reasonable to assume that a person would change the way they speak depending on whether they trust their interlocutor or not. To date, very little research has been done in this area, but there are some indications that such an effect exists. Waber et al.~\cite{waber2015} studied paralinguistic aspects in medical conversations between nurses and found that the emphasis used by an outgoing nurse when talking to an incoming nurse was significantly related to the degree of trust that the outgoing nurse reported to have on their colleagues. Further, Elkins and Derrick \cite{elkins2013} found that the variations of pitch in time were related to the degree of trust of the speaker during human-computer interactions in the form of interviews. 

In this paper, we present a preliminary set of results that we hope will contribute to answer the question of whether trust can be detected from the trustee's voice.  The experiments are done using a dataset recently collected for this purpose, the \textit{Trust-UBA Database} \cite{gauder2020}, in which subjects interact with a virtual assistant (VA) in order to respond a series of factual questions. Before each series of questions, the subject is told that the particular VA they are using was rated with a high or a low score by previous users. This initial bias is reinforced during the task by having the VA respond all or only some of the questions correctly, respectively. In this paper, we investigate whether it is possible to automatically detect which type of VA the user is interacting with, a reliable or an unreliable one. 

To this end, we implemented a classification system based on a set of features extracted automatically from the user's speech. The features were motivated by the work done on speech directed to \emph{at-risk} listeners like infants, non-native speakers and people with hearing impairment, which we believed would share similar characteristics to speech directed to an unreliable VA. Research shows~\cite{hazan2015,scarborough2007,uther2007,saint2013motherese} that non-native- and infant-directed speech include some of the following characteristics: vowel hyper-articulation, a decrease in speech rate, an increase in the number and length of pauses, and an increase in pitch excursions. Further, work on speech directed to computers that make mistakes have been found to have similar characteristics \cite{oviatt1998}. Based on these works, we designed a set of features aimed at capturing these effects.

Our experiments on the Trust-UBA database using these features show that it is possible to detect whether a user is talking to a reliable or an unreliable VA based on their speech with an accuracy up to 76\%. Note that we are not directly detecting mistrust but rather, a proxy given by the reliability of the VA. Yet, we have evidence that indicates that users did trust the unreliable VA less than the reliable VA~\cite{gauder2019,gauder2020}. These findings suggest that we may indeed be able to detect the trust level from a user's speech, though further experiments are needed to confirm these findings on larger datasets with less controlled scenarios.

\section{Trust-UBA Database}
\label{sec:trustdb}

The Trust-UBA database is composed of human-computer speech interactions where a subject is asked to respond a series of factual questions with the help of a VA. The questions were shown in the screen, one at a time, and the subject was asked to consult the VA for the answer and write their own response and that of the VA in separate cells. Most questions are designed to make it very unlikely that subjects will know the answer.
Before each question-series the subject was told that previous users of the VA rated it with a high (4.9 out of 5) or a low score (1.4 out of 5), which we will call the H and the L conditions, respectively. The initial score was then further reinforced during the interactions by having the VA respond correctly to all questions, in the H condition, or incorrectly to some questions, in the L condition. Also, during surveys occurring after the sixth, twelfth and eighteenth question, the system reminded the subject which the initial score was, hence reinforcing the bias.

We recorded 50 sessions from 50 different subjects at our laboratory. Each session was composed of two series, one L and one H, presented in random order.
Further, 110 other subjects performed the task from their homes through the internet, but these sessions were not used for the experiments in this paper. 
The same 36 questions were asked in every session. Each question could appear either in the L or the H condition during a session, except for 6 of the questions which only appeared in the L condition. Some waveforms suffered transmission errors due to bad internet connections and arrived at the server corrupted. These waveforms were discarded for our experiments.

Finally, the speech recognizer implemented as part of the VA failed for some waveforms, causing the VA to ask the user to repeat the question. In previous works, it was shown that users change the way they speak when a system fails to understand their speech \cite{oviatt1998}. These errors, which happened randomly in both L and H conditions, could then obscure the effect given by the intended reliability of the VA. 
For this reason, for the experiments in this paper, we discarded all questions within a series that came after an unplanned system error, assuming that the effect of a system error carried over until a new series started and the user was told that this was now a new VA with a different score. Further, for each question in the series, we used only the first waveform from the user that was not a mistake which required repetition (e.g., stopping the recording before finishing the question). These waveforms had an average duration of 5 seconds.

\section{Feature Description}
\label{sec:features}

As described in the introduction, the features used in this paper are motivated by the work done on speech directed to at-risk listeners. In particular, we focused on the characteristics described in \cite{oviatt1998} as being related to hyper-articulation, which include more frequent and longer pauses, slower speech rate, clearer differentiation of vowel space with respect to formant values, and increased pitch and expansion of pitch range. The features described below aim to represent these characteristics using measures that can be automatically extracted from the waveforms. The only manual annotation done on this dataset is the transcription, which is first done automatically and then corrected manually when necessary.

In order to detect the start and end of the speech for each utterance, and the duration of intermediate pauses, forced alignments to the manual transcriptions were performed using Montreal Forced Aligner \cite{montreal}. The extracted features are computed only over regions determined as speech by the forced aligner, and pauses shorter than 50 ms are considered part of the surrounding speech. Speech duration was calculated in 2 different ways: considering the full duration $T_F$ from the start to the end of speech; and considering only the speech regions $T_S$, ignoring pauses between speech regions.

We computed a total of 16 features for each waveform, 3 related to speaking rate, 7 related to pitch, 2 related to energy, and 4 related to formant information.

\textbf{Syllable rates} including and excluding pauses, were calculated dividing the number of syllables by $T_F$ and $T_S$ respectively. The number of syllables in each utterance was calculated from the transcriptions using Syltippy,\footnote{https://github.com/nur-ag/syltippy} a Spanish syllabification tool based on \cite{hernandez2013}. \textbf{Pause to speech ratio} was also calculated as the total pause duration divided by the total speech duration $T_S$.

\textbf{Pitch features} were extracted using frame level estimates of the fundamental frequency (F0). F0 was calculated using OpenSmile's \emph{smileF0} configuration file, with an F0 tracking frequency range of 100 to 620Hz over frames of 50 ms shifted by 10 ms. OpenSmile assigns frames estimated to be unvoiced an undefined F0 value. The resulting F0 signals were further masked, turning all F0 values detected over pause regions into undefined values. The resulting signal was turned into a logarithmic scale and split into regions, defined as a sequence of consecutive frames separated by more than 50 ms of unvoiced frames. Finally, we computed the following summarized values: range, given by the difference between the 95th and 5th quantiles over all values; median over all values; mean and standard deviation over the regions of the median within each region; mean and standard deviation over the regions of the range within each region; final slope, calculated using a linear regression over the last 25 (defined) frames of the F0 signal.

\textbf{Energy features} were given by the range and the slope over the last 25 frames for the energy signal extracted using OpenSmile. As for the pitch, this signal was restricted to have undefined values over unvoiced regions and turned into a logarithmic scale before computing the features.

\textbf{Formant features} were extracted for the first two formants. The formant estimates were obtained over voiced frames using OpenSmile and divided into regions as for the F0 signal.
For each of the formants we then calculated the mean and standard deviation of the ranges over the regions.

\section{Experimental Design}
For this paper, we used only the sessions from Trust-DB recorded at the school laboratory, since they had better sound quality and much fewer transmission and system errors. To make the best use of these data, the experiments were done using a leave-one-speaker out (LOSO) strategy, where a model is evaluated for each speaker using all the other speakers for training. The scores generated for all speakers using each corresponding model were pooled together to obtain one set of scores on the full dataset. 

The experiments were performed using a subset of the speakers for which at least 12 questions were available on each of the two conditions considering, for this count, only the questions (1) without transmission errors, (2) before the first system error in the series, and (3) that are not in the list of 6 questions that only appear in the L condition. 
We considered two different tasks: 
\vspace{-0.1cm}
\begin{itemize}[leftmargin=*]
 \setlength\itemsep{0em}
    \item {\bf Question-level:} The unit of classification is each question within each series. In this case, the goal is to detect the condition (L or H) for the series in which the question is found.
    \item {\bf Series-level:} The units are the question-series. The ground truth in this case is the initial score for the series.
\end{itemize}

\subsection{Normalization}
\vspace{-0.1cm}

\label{sec:norm}
The features described in Section \ref{sec:features} are likely to be highly affected by the speaker identity. This effect could potentially be more salient in the features than the condition we aim to detect. For this reason, we normalized the question-level features described in Section \ref{sec:features} over each subject's data, after filtering out questions with transmission errors or after the first system error in the series. Since the number of questions in each condition after this filtering may be different, we computed weighted statistics rather than standard ones, so that the statistics did not reflect the imbalance in the conditions. The mean $\mu$ was computed as an average of the means for both conditions and the standard deviation was calculated as the square root of the weighted variance given by $\sum_{i=0}{N} w_i (x_i-\mu)^2$, where N is the number of questions for the subject (for both conditions), $x_i$ is the feature value for question $i$ and the weights $w_i$ are given by $1/(2n_{c_i})$, where $c_i$ is the condition of the question $i$ and $n_{c_i}$ is the number of questions for this condition available for the subject. The normalization is finally done by subtracting from each feature the weighted mean and dividing by the weighted standard deviation for that feature.

Beside the user, another factor that could affect the features is the type of question. Some questions were simple to formulate (e.g., \textit{``Define the word bank''}), some were more complex or longer (e.g., \textit{``In what year did Hungary join the European Union?''}) and some were composed of two clear parts (e.g.,\textit{``Which are the three largest countries in the world, from high to low?''}) which most speakers divided up using a pause. For this reason, we explored the option of further normalizing the features by question. That is, after features are normalized by speaker, we computed the statistics per question over those features using only the training data. Those statistics were then used to normalize both the training and the test samples. This was done separately for each model being trained and the corresponding test samples for that model. 
The question statistics were also computed using weights to ensure that they were not biased by the imbalance between conditions for that question. 
For the six questions that appeared only in the low condition, weighted statistics could not be computed, since one of the conditions does not have samples. This was not a problem, since, as we will see, these questions were eliminated by the balancing process described below.

\subsection{Balancing Conditions for each Question when Training}
\label{sec:balancing}
Since each of the 36 questions used for the experiments did not appear the same number of times in the L and H condition, we undersampled each of the questions across the training data for each model to obtain exactly the same number of L and H cases. At this stage, the questions that appeared only in the L condition were discarded.
This balancing of condition per question avoids a possibly optimistic result where the model could be using the features to identify the questions, since the identity of the question would contain information about the condition. The undersampling is done randomly using 10 different seeds. For each seed, we ran LOSO and obtained a full set of scores on the test data. The scores obtained using different seeds were averaged obtaining a single score per sample in the test data.

Note that the balancing was only done during training, not during testing, since we only wanted to prevent the model from learning about the imbalance. During testing, all available questions (except the 6 that never appear in the H condition) were used to compute the summarized features, as described below.

\subsection{Summarization of Features per Series}
\label{sec:levels}
For the question-level experiments, the features input to the model were the original features described in Section \ref{sec:features}, normalized or not.
For the series-level experiments, we computed features that summarize the distribution of each feature over the series. That is, given the questions within a series, we computed the 25, 50 and 75\% quantiles for each feature. This resulted in 48 summary features per series (16x3). These summaries were used as the features for the series-level experiments. 

\subsection{Random Forest}

Classification was done using random forests \cite{hastie2001}, which are ensembles of decision trees designed to reduce the high variance of the estimators commonly found by individual decision trees.
We trained random forests consisting of 500 trees with a maximum depth of 20 using scikit-learn \cite{sklearn}. At each split, $\sqrt{N_F}$ features were randomly considered as candidates, where $N_F$ is the total number of input features. Splits were selected by minimizing the Gini impurity.

\subsection{Performance Metrics and Bootstrapping}
We reported results in terms of cross-entropy and accuracy. The cross-entropy is given by $-1/N\sum_i \log(p_i)$ where $N$ is the number of test samples and $p_i$ is the posterior given by the system to the true class for sample $i$. We normalized the cross-entropy by the value it would have on a system that always outputs a 0.5 posterior, $\log(2)$. The accuracy is obtained over hard decisions made by thresholding the posterior returned by the random forests with a threshold of 0.5.

In order to assess the uncertainty in our results, we obtained confidence intervals using  bootstrapping. We computed 1000 estimates of the cross-entropy by sampling the test scores by speaker with replacement. For each bootstrap sample, some speakers might be missing and others might be repeated several times. The scores for the units (questions or series) for each speaker are discarded or repeated accordingly, and the resulting set of scores is used to compute a new estimate of the cross-entropy. The confidence intervals shown in Figure \ref{fig:rf_results} are given by the 2.5 and 97.5 percentiles of the distribution of cross-entropy values estimated with this method.

\section{Results and discussion}

\begin{figure}[t]
    \centering
    \vspace{-1.5em}
    \includegraphics[width=0.9\linewidth]{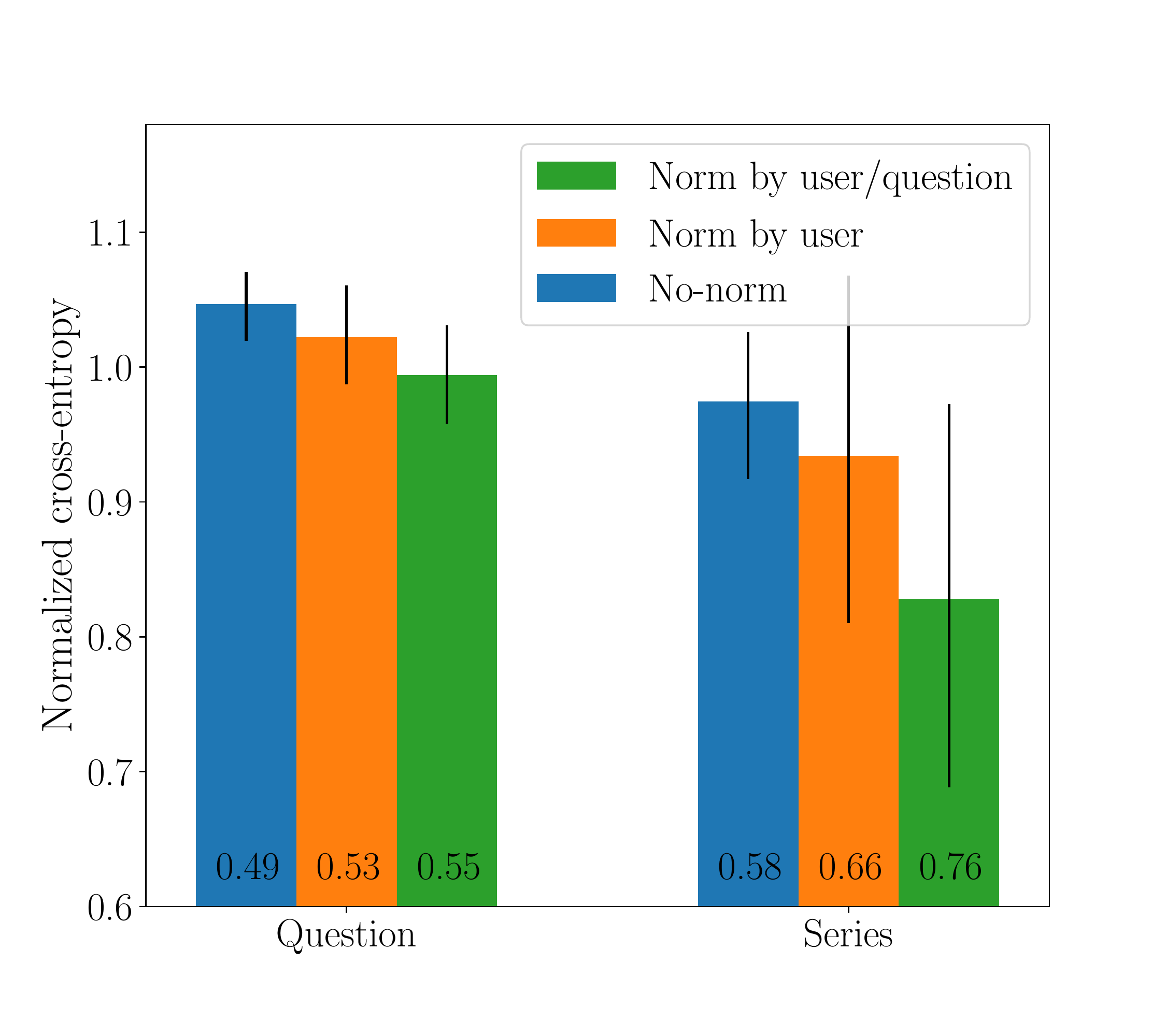}
    \vspace{-0.5cm} 
    \caption{Normalized cross-entropy obtained at question and series levels for different normalization strategies. The height of the bars indicate the normalized cross-entropy, while the error bars are the confidence intervals obtained by bootstrapping. The value inside each bar corresponds to the accuracy.}
    \vspace{-1.65em}
    \label{fig:rf_results}
\end{figure}

Figure \ref{fig:rf_results} shows the cross-entropy and accuracy for the different tasks and normalization methods. We can see that, for both tasks, normalizing by user is better than not normalizing, and normalizing by user and question leads to the highest accuracy and lowest cross-entropy. These results indicate that, as suspected, features are affected by the user and the question, and removing these effects helps the model to predict the type of VA more effectively.
Also, the performance for individual questions is close to random, even after normalization is done (which, in effect, means that the features have information about the whole session). This indicates that just a few seconds of speech are not enough to perform this complex task.

We also applied recursive feature elimination to select the best performing subset of the features described in section \ref{sec:features}. This analysis was done using the series as unit and normalizing by user and question. First, we removed each of the features one at a time and calculated the cross-entropy in the validation set. Then, the feature that when removed gave the lowest cross-entropy was excluded from the training data and the procedure was repeated until a single feature was left. Using this method we found that the best performing model needed a minimal set of 3 features: syllable rate including pauses, pitch final slope and pitch median. With this minimal feature set, the random forest model achieves a normalized cross-entropy of 0.68 with a 95\% CI [0.47,0.92] and an accuracy of 0.76. The cross-entropy is improved by 38\% by this processing, suggesting that the correlation between many of the proposed features might be degrading the model performance.  

\begin{figure}[t]
    \centering
    \includegraphics[width=1.0\linewidth]{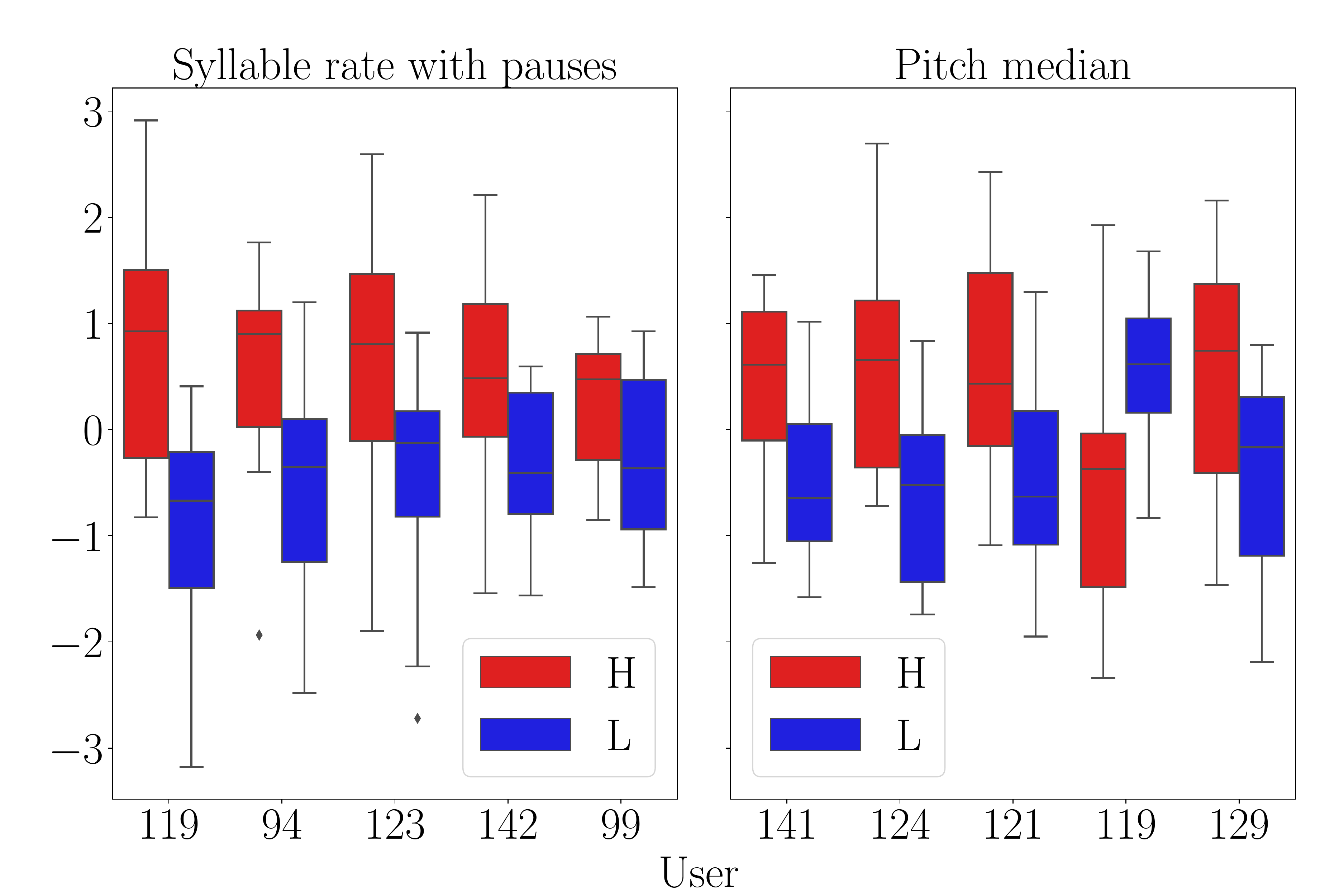}
    \vspace{-0.5cm} 
    \caption{Box and whisker plots of the syllable rate with pauses and pitch median normalized by user and question for the 5 users with highest shift of the median between low (L) and high (H) bias conditions.}
    \vspace{-1em}
    \label{fig:feats}
\end{figure}

Figure \ref{fig:feats} shows the distributions of two features selected by the recursive feature elimination method for the 5 subjects with highest absolute difference in median across conditions. It can be seen that, subjects that showed differences in syllable rate and pitch median, tended to speak faster and with higher pitch in the high bias conditions (with a single exception for the case of pitch). We also analyzed the distributions for pitch end slope but did not find any clear pattern. We hypothesize that this feature is being used by the trees to condition the use of other features depending on whether the questions are posed as a statement or as a question with the usual pitch rising at the end.

Some of the users did not show significant differences in the distribution of features between L and H conditions (not shown in Figure 2 due to lack of space). This might indicate that some users might not have been affected by the reliability of the system. In fact, we found that there is a significant correlation of 0.43 (P\textless.008) between the average posterior given by the system to the true class for a session (obtained as the average posterior for the true class for the two series in the session) and the difference in the average survey scores that the user gave the VA for the H and the L series in the session (see \cite{gauder2019,gauder2020} for details on how the users scored the VA). That is, the system appears to label correctly with more confidence the subjects that reported being more affected by the reliability of the VA. This might indicate that the subjects that did not report a difference in scores between the two series also did not change the way they spoke to the VA.

Finally, we also repeated the experiments using a subset of recordings conducted remotely on a web interface, but we could not achieve results better than random with these data. A possible explanation is that, as more system errors happened in this modality, we were left with significantly fewer sessions and questions per series, making it insufficient data to do the prediction. Another possible cause is that subjects interacting with the VA from their homes use different microphones, have various background noises, and are more distracted, consequently behaving less consistently. All of these effects are likely making the task much harder than for the sessions recorded in the laboratory.

\section{Conclusions}

We present preliminary results on the prediction of a proxy for the trust that a user has on a virtual assistant (VA) during a dialog, based on the user's speech. The proxy is given by the reliability of the VA (high or low) which correlates with the user's trust in the VA. 
Our task is, given two sets of speech waveforms from the user recorded under both conditions (high or low reliability), decide which set corresponds to each condition. We show that a system can learn to perform this task with an accuracy of up to 76\%, where a random baseline would have a 50\% accuracy. Further, we show that the features that are useful to solve this task are related to those previously found to be useful for detecting speech directed to ``at risk'' listeners like infants, non-native speakers or people with hearing loss.

We would like to emphasize that the experimental design used in this paper does not correspond to a realistic use case, since it assumes that data from both conditions are available for each user during training and testing. This setup was selected for its simplicity as a first approach for assessing whether this task could be solved automatically. The results should be only interpreted as a preliminary analysis suggesting that the proposed features do indeed contain useful information about the task. In contrast, a set of 10 expert human listeners reached very low agreement when solving this same task, highlighting the inherent difficulty of the problem \cite{gauder2020}. 

Further data collection is needed to confirm the findings in this paper in a less controlled setting. Yet, some less restrictive experiments could be designed with the Trust-UBA dataset, for example, by normalizing using only some held-out high-bias condition data for each user and dividing questions for training and testing. It is not clear, though, if the amount of data available for each speaker will be enough to obtain robust results using this approach. We will explore this option in future works.

Finally, several interesting questions arise from these experiments. Are some speakers more affected by the initial score than others? Could this behavior be predicted from their personality traits? Are some phrases more prone to contain useful information about the condition than others? We will address these questions and many others in our future research.

\section{Acknowledgements}
This material is based upon work supported by the Air Force Office of Scientific Research under award number FA9550-18-1-0026. We also gratefully acknowledge the support of NVIDIA Corporation for the donation of a Titan Xp GPU.

\bibliographystyle{IEEEtran}

\bibliography{references}

\end{document}